# Corporate system of monitoring network informational resources based on agent-based approach

D.V. Lande, V.A. Dodonov, T.V. Kovalenko

Institute of Information Recording
of the National Academy of Sciences of Ukraine

*The paper provides a agent-based model, which describes distribution of informative messages, containing links to informational resources in the Internet. The results of modeling have been confirmed by studying a real network of Twitter microblogs. The paper describes stages of building a corporate system of monitoring network informational resources, the content of which is determined by links in microblogs. The advantages of such approach are set forth.*

## Introduction

Nowadays the scope of informational resource in the web space complicate prompt extraction of information required by the user, even with the help of the most powerful search systems (Google, Baidu, Яandex etc.) [1]. Involving the opinion of a large number of people, experts in the subject area is one of the ways to tackle this problem. Such opportunity, which can be named crowdsourcing, is brought about by substantial analysis of social networks, where users "vote" for some or other news materials, making hypertext links to them. It is especially relevant for segmental subject areas, corresponding to corporate users' needs. In spite of the fact, that social networks analysis is in itself a difficult scientific task, search possibilities available in some of them bring hope to solve the described problem.

Within the framework of this research we suggest an approach to building a corporate system of monitoring network informational resources, the content of which is determined by links in social networks, particularly in Twitter microblogs network [2].

At the same time, it is necessary to analyze in details the processes, according to which the information containing links to informational resources is distributed in the networks. Modeling information distribution makes it possible to study the respective information processes, to reveal the trends, which may be used in exploring both the mechanisms of information transmission in such networks and the level of its impact upon people [3].

### Agent-based model of messages distribution

To build a agent-based information distribution model, first of all, it is necessary to form a near-reality virtual information space, inhabited by virtual agents, with which separate messages in the social network are associated, that encapsulate links to the Internet informational resources [4–5]. Separate agents are supposed to be able to [6]:

1) self-generate;
2) generate new agents by way of reposting;
3) "die" – disappear from the space of agents;
4) receive likes from other agents.

The agent has a "potential", depending upon his age, ring of authority (links to him) and fruitfulness (number of agents, reposts, generated by him). With varying four parameters of management it became possible to model the profiles of news items behavior. As a result of our research we have implemented the program of the agents' space evolution, explored the evolution of the agent-based system, found the analogies with real topical information streams. We have revealed the statistical trends, relating to the lifecycle of separate messages, distribution of which corresponds to the Weibull distribution. These models were verified by studying the real network of Twitter microblogs. The coincidence between the modeling results and the real network distribution parameters makes it possible to speak about the trend, inherent to real networks, and also about the efficacy of the model.

Modeling the dynamics of the whole informational stream starts from one agent. A new agent can appear by two ways. The first way consists in copying the existing agent with the repost operation. The agent's self-generation is also possible, which corresponds to publishing a new message. Thus at any specific time with particular probabilities any of the events can happen to each of the agents. Also at any specific time with a probability $p_s$ a new agent can appear due to self-generation.

Let us consider a life journey of one agent. The agent appears with an initial energy value $E_0$ and further on his energy changes depending upon the events, which happen to him. Les us suppose, that two events are possible: like and repost. One of these events, both of them simultaneously or none of them may happen per unit time.

Let us designate by $\varepsilon_t$ the agent's energy value at the time moment $t$. Then the energy value at the next time moment can be written down in the following way

$$\varepsilon_{t+1} = \varepsilon_t + \delta_t,$$

where $\delta_t$ is a random variable with values in {-1, 0, 1, 2}. According to the aforementioned rules of energy change, an increase in the energy by 2 corresponds to simultaneous like and repost; an increase by 1 – to repost only; the energy remains the same, if there is only like; and decreases by 1, if none of the events has happened. Therefore, we can indicate a conditional distribution $\delta_t$ at a certain energy $\varepsilon_t$:

$$P(\delta_t = 2 | \varepsilon_t = E) = p_{like}^{(E)} p_{repost}^{(E)};$$
$$P(\delta_t = 1 | \varepsilon_t = E) = \left(1 - p_{like}^{(E)}\right) p_{repost}^{(E)};$$
$$P(\delta_t = 0 | \varepsilon_t = E) = p_{like}^{(E)} \left(1 - p_{repost}^{(E)}\right);$$
$$P(\delta_t = -1 | \varepsilon_t = E) = \left(1 - p_{like}^{(E)}\right)\left(1 - p_{repost}^{(E)}\right).$$

These formulas are true for $E > 0$. Hereinafter we will use the following designation $P_\Delta^{(E)} = P(\delta = \Delta | \varepsilon = E)$. The process of agent's energy change can be considered as an integer-value random walk with transition probabilities

$$p_{ij} = \begin{cases} P_{j-i}^{(i)}, & (j-i) \in \{-1, 0, 0, 1, 2\}, \ i > 0; \\ 1, & i = j = 0; \\ 0, & \text{otherwise} \end{cases}$$

Since the energy value at the next moment of time depends only upon the energy value at the previous moment of time, the stochastic sequence $(\varepsilon_0, \varepsilon_1, ..., \varepsilon_t, ...)$ is a Markovian chain with transition probabilities $p_{ij}$.

As a result of modeling we have found, that the average lifetime and quantity of likes and reposts in this model are distributed in accordance with the Weibull probability density function [3]:

$$f(x) = \begin{cases} \dfrac{k}{\lambda}\left(\dfrac{x}{\lambda}\right)^{k-1} e^{-\left(\frac{x}{\lambda}\right)^k}, & x \geq 0; \\ 0, & x < 0. \end{cases}$$

The parameters of the Weibull distribution $k$ and $\lambda$ were obtained by the maximum likelihood method. Given the initial parameters, the obtained values are the following: $k = 1.9$, $\lambda = 3.8$.

The obtained results of modeling were compared with findings of studying the news items lifecycle in Twitter microblogs network, where, in particular, characteristics of increase in number of special reposts (retweets) of chosen messages were analyzed. The distribution of likes and retweets in this case, as well as in the model, corresponded to the standard Weibull distribution, and the parameter $k$ coincided with the model parameter to a high precision (fig. 1).

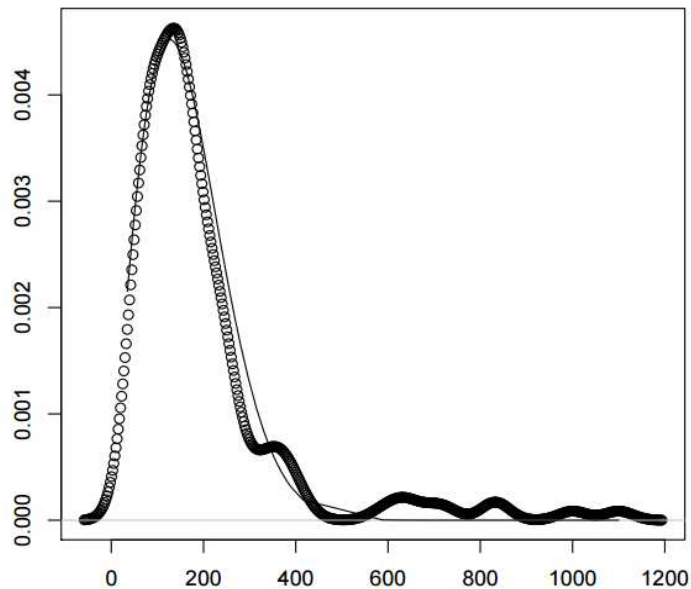

Fig. 1 – Probability density function of quantity of retweets, obtained from the real network (approximation of Weibull distribution, given $k = 1.9$, $\lambda = 180$)

## Building the database of up-to-date informational materials from the Internet

During May 2016 the authors were conducting an experiment in collecting messages from Twitter microblogs network, for which purpose the packet of 100 queries in business topics was processed in the search interface of this network on a periodic basis. As a result we obtained the following quantitative data, related to the amount of the Internet informational resources, to which there were links in Twitter-messages.

1. About 100 000 messages on elementary requests to Twitter for May 2016 were scanned.
2. 58% of messages contain hypertext links to web-resources in the Internet.
3. The number of unique hyperlinks makes up 48%.
4. The number of hyperlinks to the same sources is distributed according to the power function.
5. There are certain problems with identification of external links, connected, first of all, with use of "short links" – redirection with the following base addresses:

- http://migre.me/
- http://bit.ly/
- http://ow.ly/
- http://tinyurl.com/

- https://lnkd.in/
- https://goo.gl/
- http://wp.me/
- http://j.mp/
- http://dlvr.it/

6. The most frequently cited Internet resources:

- https://youtu.be (https://www.youtube.com)
- http://fb.me/ (https://www.facebook.com/) – public pages
- https://vk.com/
- https://twitter.com/ – links to the same social network
- https://plus.google.com/
- http://livejournal.com/

So, the practice has shown that the distribution of links to the Internet informational resources, encapsulated in social networks messages, more likely corresponds to power function (fig. 2). In accordance with this fact we have made an update of the foregoing model, that is, change of reposting probability for messages, containing links to external informational resources.

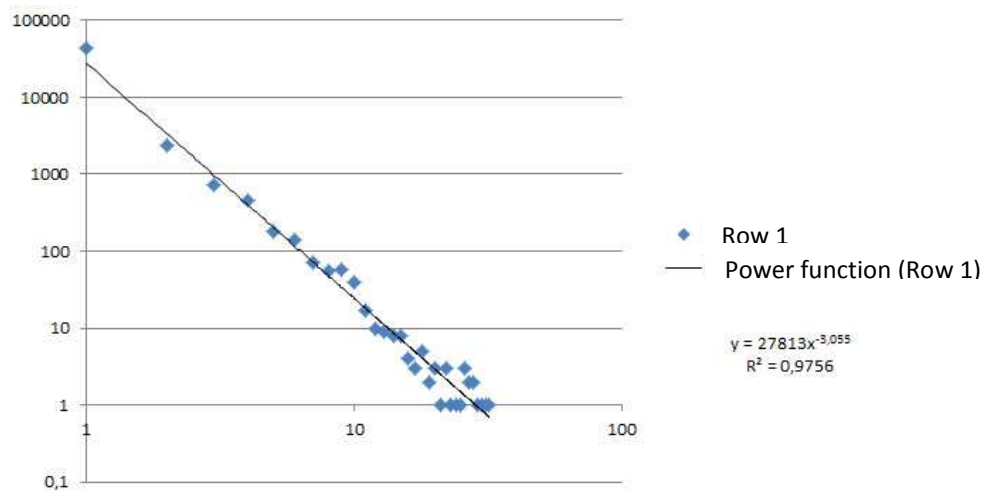

Fig. 2 – Distribution of the number of reposts of messages, containing links to external informational resources

On the grounds of the obtained information on distribution of messages, to which there are links from the microblogs network, we suggest the following "crowdsourcing" scheme of building the database of the corporate system of monitoring network informational resources (fig. 3).

This procedure is divided into the following major stages:

1. On behalf of the corporate user the queries to the microblogs network are created, for example: *Banks of Ukraine; Diamantbank; Ukrsibbank; PrivatBank; Khreshchatyk Bank; Platinum Bank; Credit Dnepr.*

2. The formed "broad" query packet is transmitted to the program, scanning the microblogs network, due to which the corporate server regularly receives messages, formally relevant to these queries.

3. Extraction of hyperlinks to the external network informational resources from the scanned messages.

4. Processing of hyperlinks, opening of "short addresses", sorting of hyperlinks, ranking of separate documents and external resources.

5. Scanning of external informational resources, corresponding to the selected hyperlinks, primary processing of obtained documents, bringing them in line with the input format of the used corporate information and analytical system.

Uploading of the formed informational stream to the corporate information and analytical system, providing corporate users with access to it.

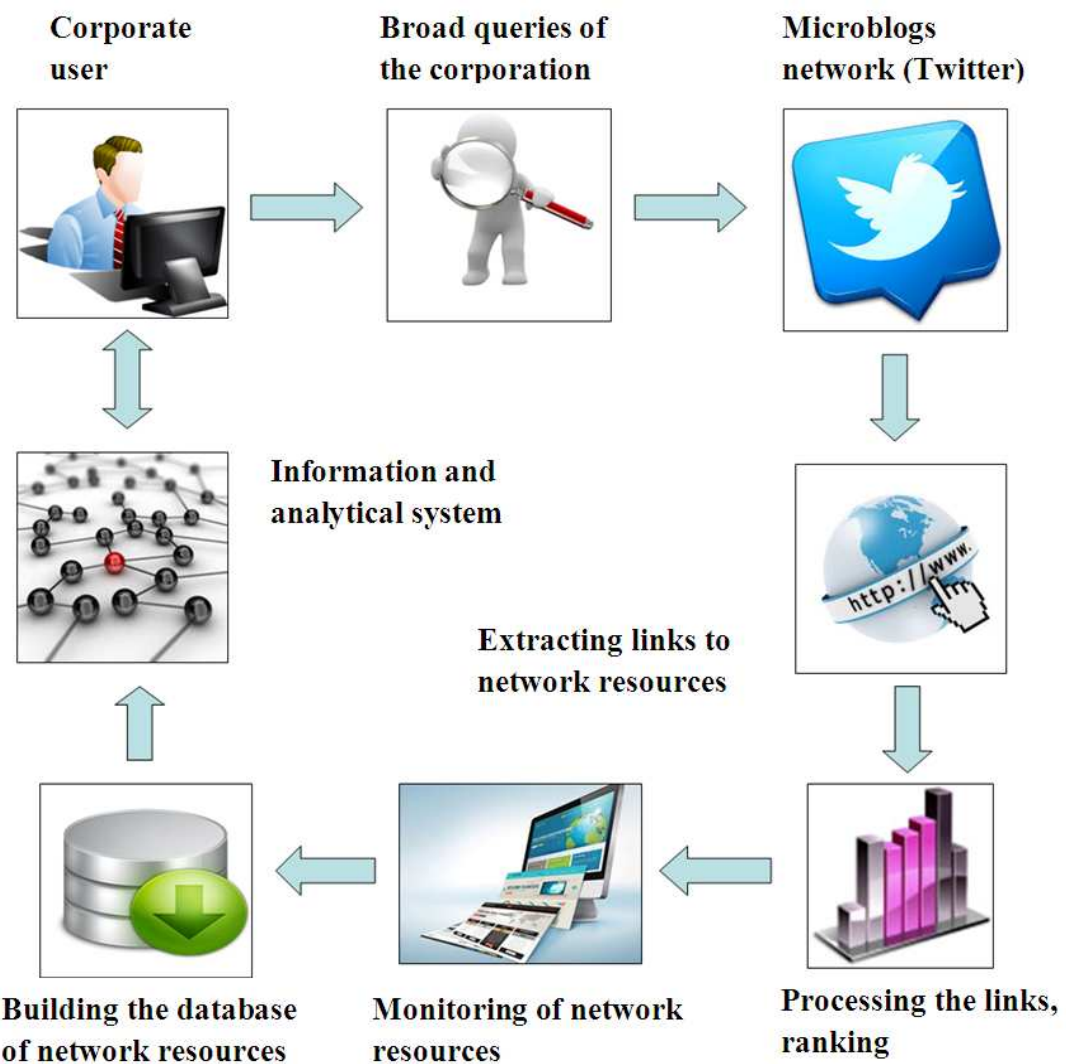

Fig. 3 – General scheme of building the database of the corporate system of monitoring network informational resources

**Conclusions**

As a result of the described research we built a agent-based model of distribution of informative messages, containing links to informational resources in the Internet. The results of modeling were verified by studying the real network of Twitter microblogs.

The revealed trends may be used in building the databases of information and analytical systems, in studying the anomalies in statistics of links to particular informational materials, and respectively, in revealing information operations, artificially supported information campaigns [7].

The described approach to forming databases on the grounds of recording the links in microblogs to informational resources, alongside with substantial reduction of information space coverage, offers the following advantages:

1. Efficiency – the informative message gets to the database of the information and analytical system in real-time mode as fast as the first user shared a link to it.

2. Coverage of the main information materials in the particular topic. Taking into consideration the opinion of interested users, whose messages content satisfies broad corporate queries. Possibility to rank information materials, proceeding from the interests of the social networks users.

3. Compactness of databases and, consequently, convenience of access for end users. Predictability of databases capacity, dynamics of their provisioning.

4. Technological compatibility with the existing information and analytical systems and content-monitoring systems.

5. Possibility to reveal information campaigns, operations.